%% file: root.tex
\documentclass[letterpaper, 10 pt, conference]{ieeeconf}  

\IEEEoverridecommandlockouts                              

\usepackage{amsmath,amsfonts,amssymb}
\usepackage{graphicx}
\usepackage{booktabs}
\input{img/sysid_beta_values} 
\usepackage{cite}
\usepackage{xcolor}
\usepackage{bm}
\usepackage[]{units}
\usepackage{mathtools}
\usepackage{textcomp}
\usepackage{wasysym} 
\usepackage{hyperref}
\usepackage{cleveref}

\newcommand\mynotes[1]{{}}

\newlength{\figcapabove}
\newlength{\figcapbelow}
\renewcommand{\c}{{\mathrm{c}}} 			

\renewcommand{\b}{{\bm{b}}}     			
\renewcommand{\phi}{\varphi} 		    
\newcommand{\current}{\bm{i}} 		    
\renewcommand{\deg}{\text{\textdegree}}   
\newcommand{\g}{{\bm{g}}}  				
\newcommand{\reg}{\textregistered\,}	
\newcommand{\SP}{\mathrm{SP}} 	 		
\newcommand{\m}{\tilde{\bm{m}}} 		
\newcommand{\n}{\bm{n}} 				
\newcommand{\M}{\bm{\mathrm{M}}}		
\newcommand{\R}{\bm{R}}		            
\newcommand{\Zero}{\bm{0}} 		 		
\newcommand{\I}{\bm{\mathrm{I}}} 		
\newcommand{\Act}{\bm{\mathcal{A}}} 	
\newcommand{\T}{\top} 			        
\newcommand{\p}{\bm{p}}                 
\newcommand{\torque}{\bm{\tau}}         
\newcommand{\force}{\bm{f}}             
\newcommand{\jacobi}{\bm{\mathrm{J}}}   
\newcommand{\skews}[1]{\mathrm{skew}\!\left\{#1\right\}}
\newcommand{\norm}[1]{\lVert #1 \rVert}

\newcommand{\tmod}{\mkern2mu\mathrm{mod}\mkern2mu}   
\newcommand{\rank}[1]{\mathrm{rank}\,\{#1\}}

\newcommand{\alloc}{\bm{\Lambda}}
\DeclareMathOperator*{\argmin}{arg\,min}

\newcommand{\Bprefix}[1]{\ensuremath{\prescript{}{\mathcal{B}}{#1}}}
\newcommand{\fW}{\mathcal{W}}                                   
\newcommand{\fH}{\mathcal{H}}                                   
\newcommand{\fB}{\mathcal{B}}                                   
\newcommand{\eH}[1]{\ensuremath{\prescript{}{\fH}{\bm{e}_{#1}}}}  
\newcommand{\eB}[1]{\ensuremath{\prescript{}{\fB}{\bm{e}_{#1}}}}  
\newcommand{\torqueB}{\Bprefix{\torque}}

\renewcommand{\r}{\bm{r}}               
\newcommand{\mAbs}{\ensuremath{\tilde{m}_0}}

\Crefname{equation}{Eq.}{Eqs.}
\Crefname{figure}{Fig.}{Figs.}
\Crefname{section}{Sec.}{Secs.}
\crefname{appendix}{App.}{Apps.}

\def\<{ \begin{bmatrix} }
\def\>{ \end{bmatrix} }

\title{\LARGE \bf
Electromagnetic Micro-Guidewire Control in Large Workspaces 
}
\author{Jasan Zughaibi$^{1,\dagger}$, Elia Jaggy$^{1,\dagger}$, Valentin Gantenbein$^{1}$, Denis von Arx$^{1}$,\\
Cristiano Sartini$^{1}$, Jonas K\"uhne$^{1}$, Oliver Brinkmann$^{1}$, Pascal Ernst$^{1}$,\\
Salvador Pan\'e$^{1}$, Quentin Boehler$^{2}$, Michael Muehlebach$^{3}$, and Bradley J. Nelson$^{1}$
\thanks{$^{\dagger}$These authors contributed equally to this work.}%
\thanks{Corresponding author: Jasan Zughaibi ({\tt\small zjasan@ethz.ch}).}%
\thanks{$^{1}$Multi-Scale Robotics Lab, ETH Zurich, 8092 Zurich, Switzerland.}%
\thanks{$^{2}$Medical Robotics Lab, ETH Zurich, 8092 Zurich, Switzerland.}%
\thanks{$^{3}$Learning and Dynamical Systems Group, Max Planck Institute for Intelligent Systems, 72076 T\"ubingen, Germany.}%
}

\begin{document}

\maketitle
\thispagestyle{empty}
\pagestyle{empty}

\begin{abstract}
Electromagnetic navigation requires sufficient actuation at clinically relevant distances due to limited magnetic volumes and coil currents. We combine real-time pose feedback with constrained convex optimization, dynamic feedback, and repetitive control to achieve energy-efficient micro-guidewire steering inside realistic anatomical models. Experiments with a clinically oriented, three-coil electromagnetic navigation system and a 0.6 mm-diameter tip magnet demonstrate angular tracking with root-mean-square errors below 0.25 degrees at distances up to 55 cm from the coil cover. Nullspace current redistribution maintains accurate tracking under active 45 A coil-current constraints. Compared with conventional field alignment, we demonstrate that pose-dependent torque-based allocation substantially reduces current demand, with the efficiency benefit retained at a pose-feedback rate of 15 Hz. These results demonstrate how real-time state information and optimization can extend electromagnetic guidewire control toward clinically relevant working distances.
\end{abstract}

\section{Introduction}
\label{sec:Introduction}

Electromagnetic navigation is rapidly emerging as a key technology in medical robotics, with applications spanning minimally invasive surgery and targeted drug delivery \cite{landers2025clinically}. It enables wireless control over magnetic tools that range from micro- and nanorobots to continuum devices such as catheters and guidewires \cite{petruska2020magMethodsRobot, gao2017micronano, zhao2022tele_mag, mesot2024teleoperated}. A central challenge is generating sufficient actuation over clinically relevant distances, particularly for micro-guidewires which have small magnetic volumes~\cite{pancaldi2020flow}. Magnetic fields decay approximately with the inverse cube of distance, while current and thermal limits constrain the field strengths attainable with an electromagnetic navigation system (eMNS). Crucially, however, the effective workspace of an electromagnetic navigation system (eMNS) is not determined by its hardware alone. Recent work showed that real-time pose feedback enables energy-efficient current allocation, expanding the operational workspace of eMNS by up to an order of magnitude relative to conventional (open-loop) field alignment control \cite{zughaibi2025workspace}. In this work, we translate this feedback methodology to magnetic micro-guidewire navigation in anatomically realistic replicas of human vasculature.

\begin{figure}[t]
    \centering
    \includegraphics[width=\columnwidth,trim={89.28pt 290.07pt 34.52pt 239.56pt},clip]{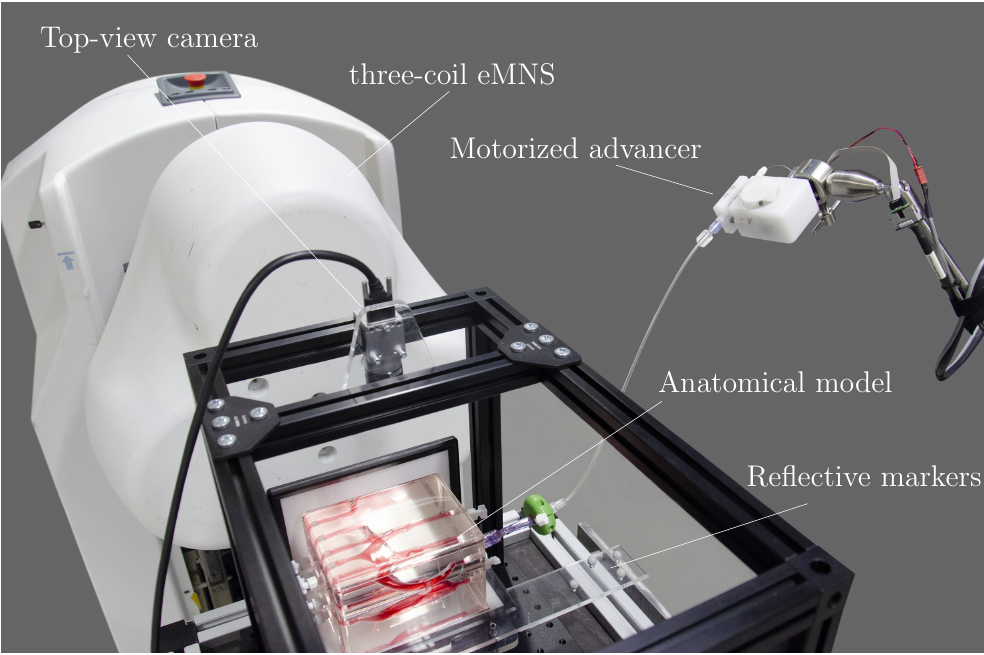}\\
    \includegraphics[width=\columnwidth,trim={0 10pt 0 5pt},clip]{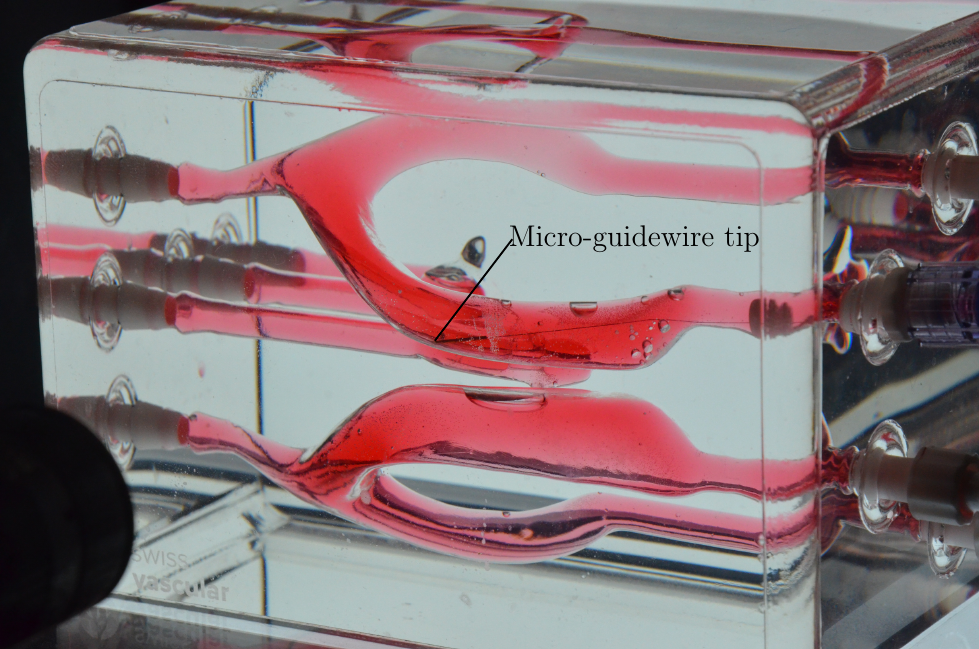}%
    \vspace{\figcapabove}
    \caption{Experimental setup. Top: \mynotes{Navion}three-coil eMNS with the top-view camera, the motorized advancer, the anatomical model, and the reflective markers. Bottom: close-up of the patient-specific in vitro vascular model (Swiss Vascular \textregistered) with the magnetic tip of the micro-guidewire inside a ventricle. The model is filled with water dyed red solely to enhance visual contrast.
}
    \label{fig:setup}
    \vspace{\figcapbelow}
\end{figure}

In conventional field-alignment control, the operator specifies the magnetic-field direction, and the guidewire is steered by the tendency of its magnetic tip to align with the externally applied field. Although intuitive, this approach does not exploit the nonuniqueness of magnetic torque generation: fields that differ in their component parallel to the tip’s magnetic moment produce identical torques, yet can require substantially different electrical effort. Exploiting this redundancy to select the minimum-current realization throughout navigation requires real-time pose feedback, because the current-to-torque mapping depends on the tip position and orientation. This enables the same task to be realized with a fraction of the energy \cite{zughaibi2025workspace}. Beyond reducing energy consumption, this efficiency gain introduces an additional design degree of freedom: a given eMNS can operate over a larger workspace or actuate objects with smaller magnetic volumes within the same hardware limits. Alternatively, a given navigation task could be performed with a more compact eMNS, potentially enhancing its economic viability.

Micro-guidewires provide a clinically important setting in which to exploit this principle. Conventional guidewires are steered by hand: the clinician advances, retracts, and rotates the proximal end while relying on a pre-shaped distal tip to select the desired vascular branch. Decades of refinement have established this simple, inexpensive, and effective approach as the clinical standard. However, friction along tortuous pathways complicates transmission of proximal steering inputs, and the pre-shaped tip cannot be actively reshaped after insertion. Magnetic guidewires apply steering torques directly at the distal end through an embedded magnet or magnetized material, bypassing torsional transmission along the shaft and enabling remotely controlled tip orientation during advancement~\cite{zhao2022tele_mag,mesot2024teleoperated}. Computer-controlled magnetic actuation further enables teleoperation from a radiation-shielded control room or a remote site and provides a foundation for precise tip positioning, active rejection of physiological disturbances, and partial automation of navigation tasks \cite{heemeyer2025teleSR}.

Beyond efficient torque generation, coil-based eMNS provide high actuation bandwidth through rapid modulation of coil currents, without mechanically repositioning permanent magnets \cite{stereotaxis2006align, valdastri2019levitation, kladko2024magnetosurgery, hongsoo2020vascular}. Permanent-magnet platforms have demonstrated guidewire navigation, including Stereotaxis systems with magnets weighing several hundred kilograms~\cite{hertting2005niobe} and smaller robot-mounted systems for submillimeter guidewires~\cite{zhao2022tele_mag}. However, their responsiveness is constrained by the inertia of the magnets and their positioning mechanisms, while rapid mechanical motion near the patient introduces additional safety considerations. Electrically modulated eMNS avoid these mechanical constraints and support dynamic stabilization without moving the field sources. This capability has enabled highly dynamic and complex feedback-control tasks, including on clinical-grade systems \cite{zughaibi2024balancing,singh2026levitation, sydora2026learning}.

In this work, we investigate how real-time pose feedback enables energy-efficient magnetic micro-guidewire navigation and improves tip-tracking performance. In our \textit{in-vitro} experiments, these estimates are obtained using stereo vision. We show that the energy-efficiency benefit is largely insensitive to the pose-feedback sampling rate and is retained at \unit[15]{Hz}, representative of clinical fluoroscopy~\cite{sadamatsu2016lowframe}. This finding motivates future work on fluoroscopic pose reconstruction and registration to the eMNS, with the prospect of using clinical imaging for both anatomical guidance and energy-efficient feedback control. Furthermore, recent advances in localization using passive inductive pick-up coils offer a promising route toward continuous, non-ionizing pose feedback during magnetic actuation~\cite{denis2025pickupcoil}. Their fine-wire construction and absence of active tip electronics could facilitate miniaturization and integration into micro-guidewires.

We demonstrate accurate closed-loop guidewire-tip control with a clinically-oriented, three-coil eMNS using a \unit[0.6]{mm}-diameter tip magnet with a magnetic volume of \unit[2.47]{mm$^3$}. With currents limited to \unit[45]{A} per coil, we achieve vertical-plane ($\alpha$) tracking at \unit[43.7]{cm} and horizontal-plane ($\beta$) tracking at \unit[54.9]{cm} from the coil cover, with angular tracking RMSE below $0.25\deg$ despite active current constraints. To the best of our knowledge, this combination of working distance, tip-magnet size, and angular tracking accuracy has not previously been demonstrated in closed-loop electromagnetic guidewire control. At the core of our framework is a pose-dependent convex torque-allocation scheme that extends the pseudoinverse-based approach in~\cite{zughaibi2025workspace} by exploiting nullspace redundancy to redistribute coil loading under individual current constraints. We integrate this allocation with dynamic feedback control and use random-phase multisine identification to characterize the coupled eMNS-guidewire dynamics and nonlinear distortion. We quantify the actuation-efficiency benefit against field-alignment benchmarks and demonstrate that an iterative learning control scheme compensate repeatable tracking errors. Together, these results demonstrate how energy-efficient allocation and dynamic feedback enable accurate guidewire control at clinically relevant distances, while learning-based compensation provides a foundation for future automatic rejection of periodic physiological disturbances.

The remainder of this paper is structured as follows: \Cref{sec:Experimental Setup} introduces the experimental platform, and \Cref{sec:Modeling} describes the guidewire dynamics and magnetic actuation model. \Cref{sec:optimization} presents constrained current allocation and its experimental evaluation. \Cref{sec:Control} describes the control strategies and field-alignment benchmark. Finally, \Cref{sec:Conclusion} discusses the findings and perspectives for clinical translation.





\section{Experimental Setup}
\label{sec:Experimental Setup}

\subsection{Micro-Guidewire}

The guidewire comprises a superelastic nickel--titanium (NiTi) base wire of \unit[100]{$\mu$m} diameter with a tubular permanent magnet mounted around the wire at its distal tip. The magnet has a nominal outer diameter of \unit[0.024]{in} (approximately \unit[0.6]{mm}) and a length of \unit[9]{mm}. Accounting for the hollow cylindrical geometry, the volume of magnetic material is $V = \unit[2.47]{mm^3}$. Vibrating sample magnetometry (VSM) of an identical magnet along its longitudinal axis yields an estimated remanent flux density of $b_r = \unit[0.28]{T}$. The corresponding magnetic dipole moment magnitude is $\mAbs = b_r V/\mu_0 \approx \unit[5.5 \times 10^{-4}]{A\,m^2}$, where $\mu_0$ denotes the vacuum permeability. The measured longitudinal intrinsic coercivity is $H_{\mathrm{c}} = \unit[72]{kA\,m^{-1}}$, defined as the reverse-field magnitude at which the longitudinal magnetization vanishes on the hysteresis loop. In this work, the applied magnetic flux density at the tip typically remains below \unit[10]{mT}, substantially smaller than $\mu_0 H_{\mathrm{c}} \approx \unit[90.5]{mT}$. Under these operating conditions, field-induced changes in magnetization are neglected, and the dipole moment is assumed constant in the magnet-fixed frame.

\subsection{Tracking, Motorized Advancer, and Anatomical Model}

Guidewire-tip pose is estimated using two synchronously triggered monochrome cameras (Basler\reg acA1440-2) with approximately orthogonal views and LED backlighting to enhance contrast. Background subtraction segments the magnet in each image, and principal component analysis identifies its projected longitudinal axis and endpoints. Stereo triangulation reconstructs the magnet-center position and longitudinal-axis direction in the guidewire-holder frame.

Reflective markers attached to the setup are tracked by a motion-capture system to register the guidewire-holder frame to the eMNS coordinate system, allowing the reconstructed position and orientation to be used for magnetic actuation. Images are acquired at either \unit[100]{Hz} or \unit[15]{Hz}; the latter is used to evaluate feedback at a sampling rate representative of clinical fluoroscopy~\cite{sadamatsu2016lowframe}. The \unit[0.6]{mm} magnet diameter was selected in part to facilitate optical localization.

A custom-designed motorized advancer provides axial insertion and retraction. The operator commands the motor angular velocity through a joystick, thereby adjusting the direction and speed of guidewire advancement, while magnetic actuation controls the distal-tip orientation.

We demonstrate guidewire steering in a patient-specific anatomical model of the cerebral ventricular system (Swiss Vascular\textregistered, as shown in \Cref{fig:setup}).

\subsection{Electromagnetic Navigation System}

All experiments are conducted using\mynotes{ the Navion,} a clinically oriented eMNS comprising three lateral coils arranged in a triangular configuration\mynotes{~\cite{simone2024navion}}, as shown in \Cref{fig:setup}. The coils are driven independently by commercial current-control electronics (Gold Drum HV, Elmo Motion Control\textregistered), with the commanded current vector denoted by $\current\in\mathbb{R}^3$.

An important characteristic of an eMNS for dynamic feedback control is its electrical actuation bandwidth, which determines the frequency range over which magnetic torques and forces can be effectively generated. We define the bandwidth as the frequency at which the magnitude of the transfer function from commanded to measured current falls \unit[-3]{dB} below its low-frequency value, corresponding to an amplitude reduction of approximately \unit[30]{\%}. The \mynotes{ the Navion,}electrical bandwidth was \mynotes{previously } identified as \unit[24.5]{Hz} for a current-reference amplitude of \unit[5]{A}\mynotes{~\cite{zughaibi2025workspace}}. The underlying system-identification approach is outlined in \Cref{sec:Identification of the Guidewire Dynamics}. The hardware supports currents of up to $\bar{\imath}=$\unit[45]{A} per coil, with current references updated at \unit[100]{Hz}.

\section{Modeling}
\label{sec:Modeling}
We characterize the coupled eMNS-guidewire dynamics and formulate the magnetic actuation model relating coil currents to effective tip torque. The identified dynamics inform the assessment of achievable feedback performance, while the actuation model provides the basis for pose-dependent current allocation.

\subsection{Identification of the Guidewire Dynamics}
\label{sec:Identification of the Guidewire Dynamics}

The electrical bandwidth of the eMNS alone does not determine the
achievable performance of guidewire control. The guidewire’s intrinsic mechanical dynamics are governed by its inertia, damping, and elasticity, while interactions with the surrounding fluid introduce additional viscous damping.
We therefore identify the combined eMNS-guidewire dynamics from the
commanded magnetic torque $\tau_{\phi,\SP}$ to the measured tip angle
$\phi\in\{\alpha,\beta\}$, both in air and in water. Our primary objective is to assess the dynamic limitations on
achievable closed-loop performance. The identified frequency responses provide a basis for assessing the frequency range over which effective trajectory tracking and disturbance rejection are feasible. The identification procedure additionally provides a
frequency-resolved measure of deviations from linear behavior.

We follow the frequency-domain identification procedure detailed
in~\cite{hao2022learning, zughaibi2024balancing,
2012sysID_freqDomainAppr}. The excitation signals are periodic
multisines, obtained by superimposing sinusoids of equal amplitude
at selected frequencies within \unit[0-10]{Hz}. We generate ten
realizations by independently randomizing the sinusoidal phases.
The resulting signals therefore have different time-domain waveforms but
have identical amplitude spectra. Each realization is applied for ten consecutive periods in order to increase the signal-to-noise ratio.
The first four periods are discarded to suppress transients,
and the remaining six are used for analysis. Averaging the
frequency-response estimates across realizations yields an
estimate of the best linear approximation for the
prescribed excitation. 

Repeating an identical signal estimates the variability caused
by stochastic noise and other non-repeatable disturbances.
Changing its phases additionally reveals nonlinear behavior:
an ideal linear time-invariant system would yield the same
frequency response for every realization.
The estimated standard error $\hat{\sigma}_{\mathrm{n}}$
quantifies the stochastic-noise contribution to the averaged
frequency response, whereas $\hat{\sigma}_{\mathrm{nl}}$
quantifies its total uncertainty, estimated from variability
across phase realizations. Since the latter also includes
residual stochastic noise, a substantial excess of
$\hat{\sigma}_{\mathrm{nl}}$ over $\hat{\sigma}_{\mathrm{n}}$
indicates nonlinear distortion.

\begin{figure}[t]
    \centering
    \includegraphics[width=\columnwidth]{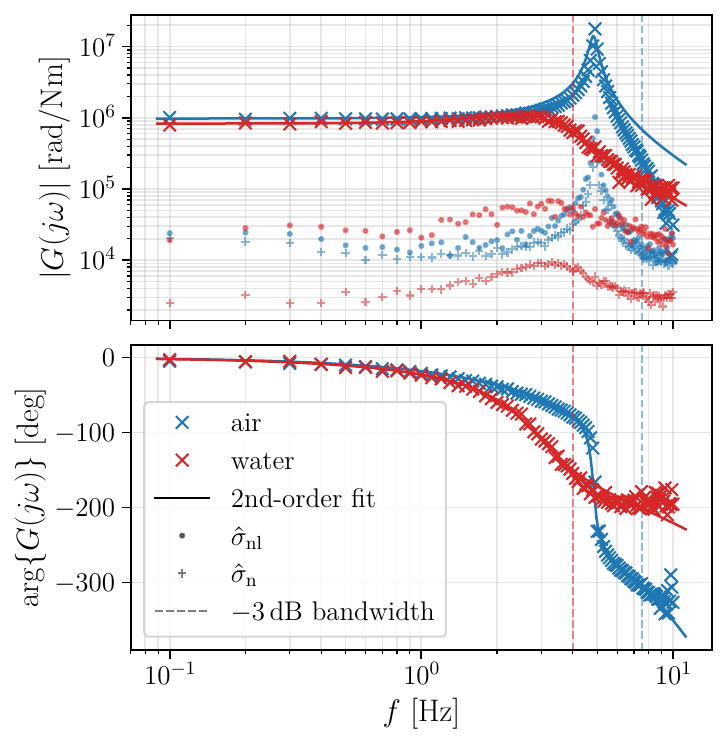}%
    \vspace{\figcapabove}
    \caption{Frequency response from commanded magnetic torque to
    guidewire-tip deflection about the $\beta$ axis in air and water.
    Crosses show the BLA estimates; dots and plus signs indicate
    the total standard error $\hat{\sigma}_{\mathrm{nl}}$ and its
    stochastic-noise contribution $\hat{\sigma}_{\mathrm{n}}$,
    respectively. Solid lines show second-order fits with delay;
    dashed vertical lines mark the \unit[-3]{dB} bandwidths relative to
    the static gain. Identified parameters, given as
    \textit{(air, water)} pairs, are
    $K=(\sidKstatAir,\,\sidKstatWater)\times10^{5}\,
    \mathrm{rad}/(\mathrm{N\,m})$,
    $f_0=\omega_0/(2\pi)
    =(\sidfZeroAir,\,\sidfZeroWater)\,\mathrm{Hz}$,
    $D=(\sidDAir,\,\sidDWater)$, and
    $T_d=(\sidTdAir,\,\sidTdWater)\,\mathrm{ms}$. The identified dynamics, particularly the static gain $K$, depend on the effective lever arm and may therefore vary with the guidewire configuration. }
    \label{fig:sysid_beta}
    \vspace{\figcapbelow}
\end{figure}

\Cref{fig:sysid_beta} shows the identified frequency
responses for the $\beta$ direction in air and in water. For each environment, we approximate the identified response by a second-order model with an effective delay,
\begin{equation}
    G_{\phi}(s)
    =
    e^{-s T_\mathrm{d}}
    \frac{K\omega_0^2}
    {s^2+2D\omega_0s+\omega_0^2},
    \label{eq:guidewire_identified_model}
\end{equation}
where $K$ denotes the static angular gain, $\omega_0$ the
undamped natural frequency, $D$ the damping ratio, and $T_\mathrm{d}$ the
effective delay. The fitted parameters for the $\beta$ direction are reported
in the caption of \Cref{fig:sysid_beta}. The combined system exhibits a \unit[-3]{dB} bandwidth of approximately \unit[4]{Hz} in water. This result highlights
the dynamic actuation capabilities of the eMNS--guidewire
system and suggests potential for active compensation of
physiological disturbances during vascular interventions.

\subsection{Magnetic Actuation Model}
\paragraph{Current-to-field mapping}
We assume that, at any position $\p\in\mathbb{R}^{3}$ within the
workspace, the magnetic field $\b \in\mathbb{R}^{3}$ and its spatial gradient $\g \coloneqq \< \partial b_x/\partial x & \partial b_x/\partial y & \partial b_x/\partial z & \partial b_y/\partial y & \partial b_y/\partial z \>^\T$ depend
linearly on the coil currents $\current\in\mathbb{R}^{3}$:
\begin{align}
    \< \b \\ \g \>
    =
    \< \Act_{\b}(\p) \\ \Act_{\g}(\p) \>\current
    =
    \Act(\p)\current.
    \label{eq:current_to_field}
\end{align}

Under the quasistatic approximation and in the absence of free currents, Maxwell's equations yield $\nabla\cdot\b=0$ and $\nabla\times\b=0$. Hence, $\nabla\b$ is symmetric and trace-free, reducing its number of independent components from nine to five. The position dependence of $\Act(\p)$ is described using a multipole expansion truncated at quadrupole order~\cite{petruska2017multipole}. Although the field model is nonlinear in position, it assumes a linear dependence on the coil currents. This approximation is sufficiently accurate for current allocation in the operating regime of the deployed \mynotes{Navion }eMNS, although deviations from linear superposition may occur at amplitudes $>$\unit[40]{A}. The model parameters are obtained using the publicly available calibration data and open-source identification procedure accompanying \cite{bernardes2026structured}. The resulting map $\Act(\p)$ is evaluated online using the measured position of the guidewire tip.

\paragraph{Magnetic wrench and effective tip torque}
The torque and force acting on a magnetic dipole moment $\m$ in a spatially varying external magnetic field $\b$ are given by $ \torque = \m\times\b$ and $\force = (\m\cdot\nabla)\b,$
which can equivalently be written in matrix form as
\begin{align}
\< \torque \\ \force \>
&=
\<
\M_{\b}(\m) & \Zero \\
\Zero & \M_{\g}(\m)
\>
\< \b \\ \g \> 
\coloneqq
\M(\m)\< \b \\ \g \>  \label{eq:field_to_torque}
\end{align}
where $\M_{\b}(\m)=\skews{\m} \in\mathbb{R}^{3 \times 3}$ and $\M_{\g}(\m) \in\mathbb{R}^{3 \times 5}$ is the corresponding dipole-force matrix \cite{petruska2015MinimumBounds}.

\begin{figure}[t]
    \centering
    \includegraphics[width=0.85\columnwidth]{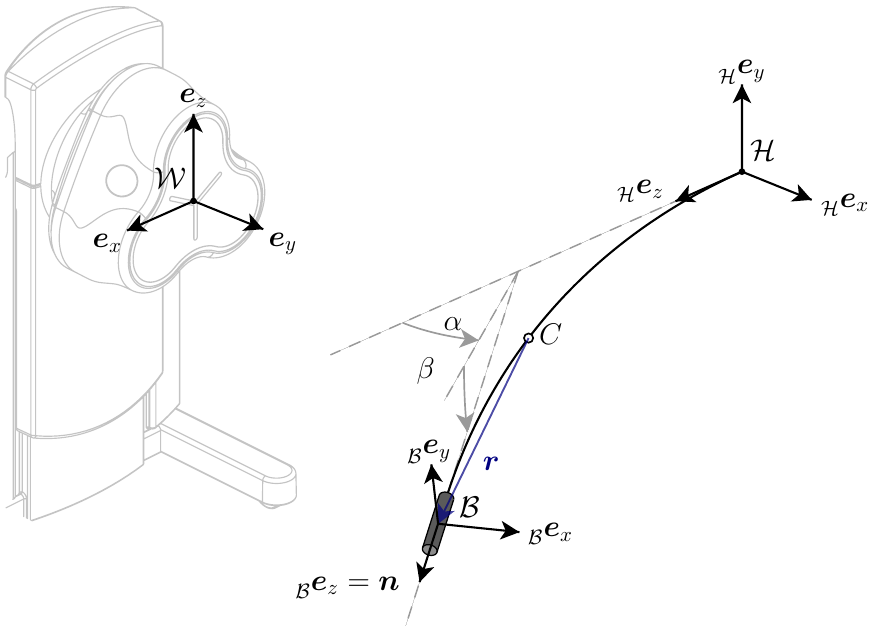}%
    \vspace{\figcapabove}
    \caption{Frames and tip-angle definitions. The world frame $\fW$ sits at the center of the coil cover. The holder frame $\fH$ is the reference frame for the tip orientation: the tip angles $\alpha$ (rotation about $\eH{y}$) and $\beta$ (rotation about $\eH{x}$) describe the direction $\n=\eB{z}$ of the dipole axis, i.e., of the body-fixed frame $\fB$ at the tip magnet; the origin of $\fH$ is drawn at the pivot point for illustration only. The lever arm $\r$ of the magnetic tip force extends from the local center of rotation $C$ to the tip magnet, where $C$ could be, e.g., a contact with the vessel wall that creates a local point of rotation; we assume $\r\approx\ell\,\n$ with a characteristic lever arm $\ell$.}
    \label{fig:guidewire_frames}
    \vspace{\figcapbelow}
\end{figure}

Axial translation of the guidewire is controlled independently by the motorized advancer stage, whereas magnetic actuation is used to control the orientation of its tip. Both the magnetic torque $\torque$ and moments induced by magnetic forces $\force$ can contribute to this orientational motion. We therefore combine these contributions into a single effective control torque $\torque_{\c}$. In particular, contact with the surrounding vasculature constrains the guidewire such that forces acting at its tip generate an additional moment about a local, geometry-dependent center of rotation $C$ (see \Cref{fig:guidewire_frames}). Rather than modeling this center explicitly, we approximate the corresponding lever-arm vector $\r$ from $C$ to the tip magnet as $\r \approx \ell\,\n,$ where $\n\coloneqq \m/\mAbs$ and $\ell$ is a constant characteristic lever arm. The effective control torque is then approximated by
\begin{align}
\torque_{\c}
&= \torque+\r\times\force 
\approx
\< \I & \ell\,\skews{\n} \>
\< \torque \\ \force \> \coloneqq\jacobi(\n) \< \torque \\ \force \> \label{eq:jacobian}
\end{align}
where $\jacobi$ denotes the wrench-to-torque mapping that accounts for the moment induced by magnetic forces about the local center of rotation. Thus, $\ell$ provides a simple approximation of how magnetic tip forces contribute to orientational control without explicitly modeling the geometry-dependent center of rotation. Expressed in the body-fixed frame $\mathcal{B}$, whose local $z$-axis is aligned with the dipole axis, the controllable torque is
$
\torqueB_{\c}
\coloneqq
\< \tau_{\c,x} & \tau_{\c,y} & 0 \>^\T,
$
where the third component is zero because neither the magnetic torque nor the force-induced moment can generate a torque about the dipole axis. The two orthogonal components, $\tau_{\c, x}$ and $\tau_{\c, y}$, are determined by the feedback controller or directly by the human operator (see \Cref{sec:Control}). The corresponding inertial-frame torque is obtained from $\torque_\c = \R^\T \torqueB_\c$, where $\R \in \mathrm{SO}(3)$ maps vectors from the inertial frame to the body-fixed frame. In the next section, we address the inverse actuation problem: determining the coil currents required to realize a given control torque $\torque_\c$ at the measured magnet pose.

\section{Convex Current Allocation under Constraints}
\label{sec:optimization}

Combining \eqref{eq:current_to_field}, \eqref{eq:field_to_torque}, and
\eqref{eq:jacobian} yields the pose-dependent linear map from coil currents
to the effective control torque,
\begin{align}
    \torque_\c
    &=
    \jacobi(\n)\,\M(\mAbs\n)\,\Act(\p)\,\current
    \coloneqq
    \alloc(\n,\p)\,\current,
    \label{eq:current_to_torque}
\end{align}
where $\m=\mAbs\n$ and $\alloc(\n,\p)$ is the allocation matrix.
By the orthogonality property of the cross product, both the direct
magnetic torque $\m\times\b$ and the force-induced moment
$\ell\,\n\times\force$ lie in the plane $\n^\perp$. Consequently,
\begin{align*}
    \n^\T\torque_\c=0,
    \qquad
    \n^\T\alloc(\n,\p)=\Zero,
    \qquad
    \rank{\alloc(\n,\p)}\leq2.
\end{align*}
For the three-coil system, the rank-nullity theorem therefore guarantees a nontrivial right nullspace. At regular poses, $\rank{\alloc(\n,\p)}=2$, so the allocation spans the complete torque plane $\n^\perp$ and its nullspace is one-dimensional. A further rank loss constitutes a magnetic actuation singularity, at which torques in $\n^\perp$ can no longer be generated independently \cite{petruska2015MinimumBounds}.

Direct allocation of desired magnetic torques and forces to coil currents using pose-dependent pseudoinverses has previously been employed for permanent-magnet levitation and microrobot actuation~\cite{berkelman2013levitation,diller2014six}. For a commanded torque $\torque_\c\in\operatorname{image}\{\alloc(\n,\p)\}$, the Moore-Penrose solution $\current_{\mathrm{pinv}}=\alloc(\n,\p)^\dagger\torque_\c$ minimizes $\norm{\current}_2^2$ among all current vectors satisfying \eqref{eq:current_to_torque} exactly, in the absence of current bounds. If the requested currents exceed the hardware limits, current saturation can prevent exact realization of the commanded torque in \eqref{eq:current_to_torque}. At each control update, we recompute the allocation from the measured magnet
pose and check a posteriori whether the current bounds are satisfied, applying $\current^\star=\current_{\mathrm{pinv}}$ only when $\norm{\current_{\mathrm{pinv}}}_\infty\leq\bar{\imath}$.

Constrained optimization, on the other hand, has been used for minimum-current synthesis of desired \textit{magnetic fields} under coil-current bounds in~\cite{lee2024constrField}. However, such \textit{field-centric} objectives do not generally minimize the current required for the underlying mechanical task. \textit{Motion-centric} allocation instead directly targets the torques and forces that govern the mechanical response. This distinction is analyzed in~\cite{zughaibi2025workspace}, where direct torque allocation enabled an order-of-magnitude improvement in energy efficiency relative to conventional field-alignment control. Constrained torque/force allocation accounting for amplifier, power-supply, and temperature limits has also been formulated using nonlinear optimization and evaluated in simulation~\cite{abbott2020powerTemp}.

\begin{figure}[t]
    \centering
    \includegraphics[width=\columnwidth]{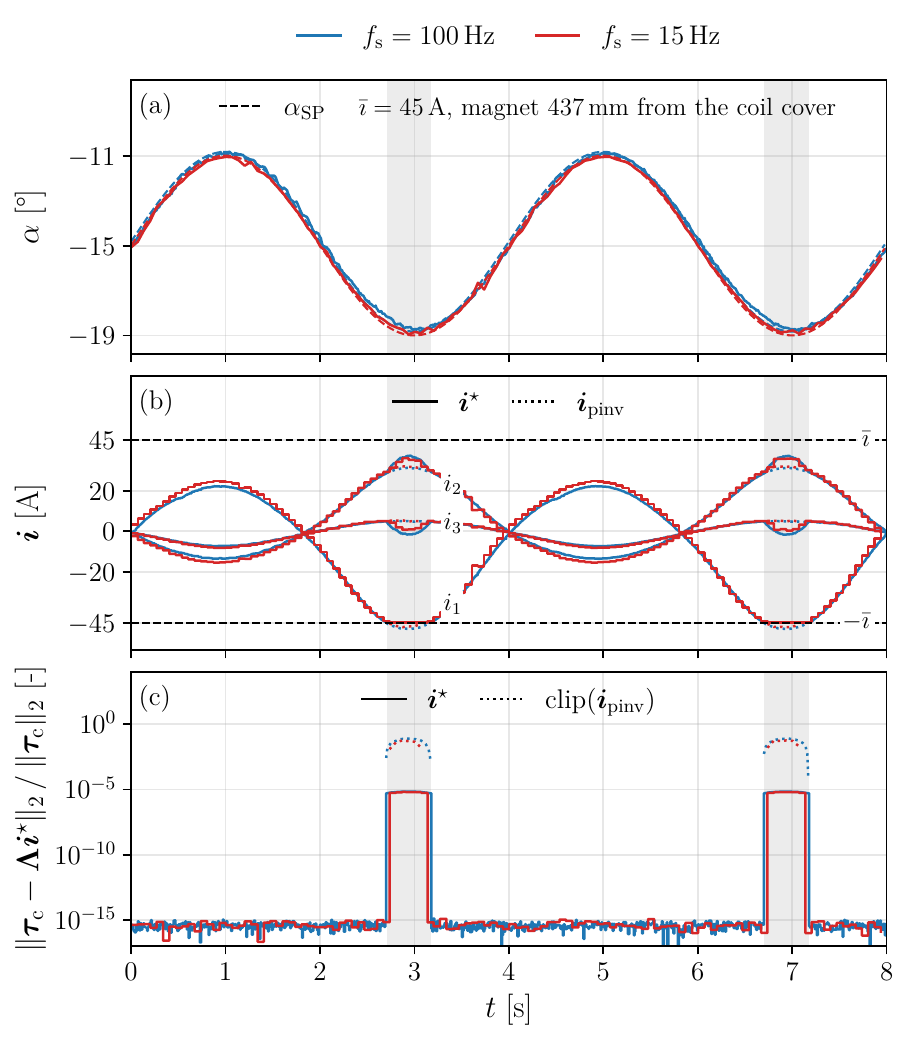}%
    \vspace{\figcapabove}
    \caption{Constrained torque allocation during tip-angle tracking in a fluid-filled vascular model, with $\bar{\imath}=\unit[45]{A}$, the magnet \unit[437]{mm} from the coil cover, and sampling rates of \unit[100]{Hz} and \unit[15]{Hz}. (a) Measured tip angle $\alpha$ and sinusoidal reference $\alpha_\SP$ (dashed). (b) Allocated currents $\current^\star$ and pseudoinverse currents $\current_{\mathrm{pinv}}$ computed a posteriori, shown where they violate a bound (dotted). Optimization holds the constrained current at $-\bar{\imath}$ and redistributes the remaining currents. (c) Relative torque residual near machine precision for feasible pseudoinverse allocation and approximately $7\cdot10^{-6}$ under active constraints, compared with \unit[6]{\%}-\unit[8]{\%} for componentwise clipping of the pseudoinverse solution (dotted). Gray bands mark intervals with an active current constraint.}
    \label{fig:allocation_limit}
    \vspace{\figcapbelow}
\end{figure}

\begin{figure}[t]
    \centering
    \includegraphics[width=\columnwidth]{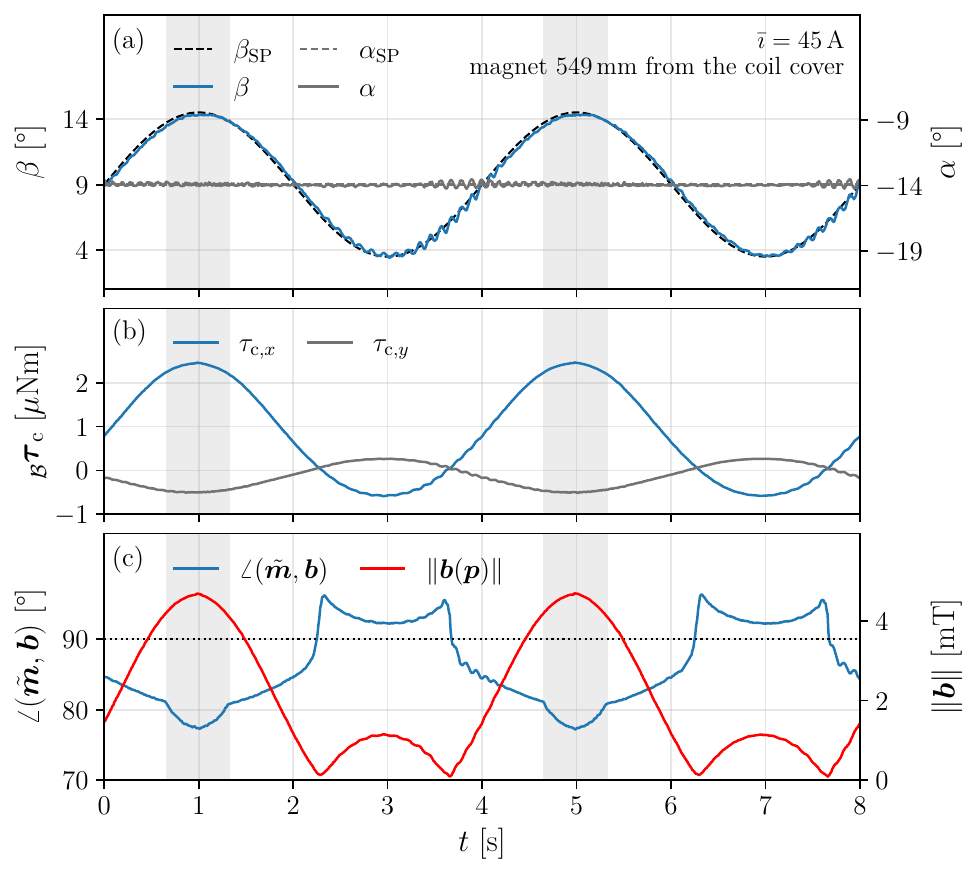}\\[4pt]
    \includegraphics[width=\columnwidth]{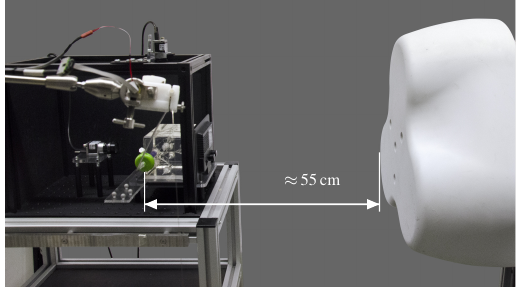}%
    \vspace{\figcapabove}
\caption{Tip-angle tracking at the largest demonstrated working distance, with the magnet \unit[549]{mm} from the coil cover and $\bar{\imath}=\unit[45]{A}$.
(a) Measured tip angles $\beta$ and $\alpha$ (right axis) with references (dashed).
(b) Body-fixed control torque $\torqueB_\c$. (c) Field--dipole angle (left axis) and field magnitude $\norm{\b}$ (right axis). The nearly perpendicular field
and force-induced moment below \unit[2]{\%} of the effective
torque yield $\norm{\torque_\c}\approx\norm{\m}\,\norm{\b}$. During active current constraints (gray bands), nullspace redistribution introduces a field component parallel to the dipole, reducing the angle to $77\deg$, while the relative torque residual remains $1.3\cdot10^{-6}$.
    Bottom: setup photograph illustrating the working distance.}
    \label{fig:max_distance}
    \vspace{\figcapbelow}
\end{figure}

Here, we incorporate coil-current bounds into the
pose-dependent torque allocation for guidewire-tip
control. By exploiting the allocation nullspace to redistribute
coil loading, we further extend the current-limited workspace
beyond that obtained with the unconstrained pseudoinverse.
Whenever the pseudoinverse solution violates a current bound,
we solve the convex regularized least-squares problem
\begin{equation}
\begin{aligned}
    \current^\star
    =\argmin_{\current}\quad &
    \norm{\alloc\current-\torque_\c}_2^2 +\epsilon\norm{\alloc}_2^2\norm{\current}_2^2
    \\
    \text{s.t.}\quad &
    \norm{\current}_\infty\leq\bar{\imath},
\end{aligned}
\label{eq:regularized_allocation}
\end{equation}
where we set $\epsilon=10^{-6}$. Scaling the current penalty by $\norm{\alloc(\n,\p)}_2^2$ ensures that both terms in the cost function have the same units. 
The problem is always feasible and strictly convex for $\alloc(\n,\p)\neq\Zero$, and is solved in real-time using OSQP~\cite{stellato2020osqp}. As $\epsilon\to0^+$, its solution converges to the minimum-$2$-norm current vector among all vectors minimizing
the torque error subject to the current bounds. This limit satisfies \eqref{eq:current_to_torque} exactly whenever that equation admits a solution with $\norm{\current}_\infty\leq\bar{\imath}$. Finite regularization results in torque errors in both magnitude and direction; however, the computed \textit{relative} residuals are negligible, of order $10^{-6}$-$10^{-5}$, in the experiments reported in \Cref{fig:allocation_limit,fig:max_distance}.

The mechanism underlying this additional workspace expansion follows
from the nullspace structure. At regular poses, all exact current
realizations can be written as
\begin{align*}
    \current
    &=\current_{\mathrm{pinv}}+\eta\bm{z},
    \qquad \eta\in\mathbb{R}, \qquad  \bm{z}\in\ker\!\{\alloc(\n,\p)\}.
\end{align*}
Since $\alloc(\n,\p)\bm{z}=\Zero$, varying $\eta$ redistributes the
coil currents while preserving
$\alloc(\n,\p)\current=\torque_\c$. Moreover,
$\current_{\mathrm{pinv}}^\T\bm{z}=0$, such that
\begin{align}
    \norm{\current_{\mathrm{pinv}}+\eta\bm{z}}_2^2
    =\norm{\current_{\mathrm{pinv}}}_2^2
    +\eta^2\norm{\bm{z}}_2^2.
    \label{eq:nullspace_current_norm}
\end{align}
A nonzero nullspace contribution therefore increases the overall energy but can reduce the largest individual coil current. Because both $\alloc(\n,\p)$ and its nullspace
vary with the magnet pose, this redistribution requires real-time pose feedback.

We experimentally explored the distance limits of tip-angle tracking in the $\alpha$ and $\beta$ control planes, with the guidewire confined to a fluid-filled vascular model. \Cref{fig:allocation_limit} demonstrates current redistribution during sinusoidal tip-angle tracking with the magnet \unit[437]{mm} from the coil cover. When the pseudoinverse solution exceeds the \unit[45]{A} current limit, the optimized allocation redistributes coil loading while maintaining a relative torque residual of approximately $7\times10^{-6}$. Componentwise clipping of $\current_{\mathrm{pinv}}$ would instead produce torque errors of \unit[6]{\%}-\unit[8]{\%}. The redistribution preserves accurate setpoint tracking through intervals of active current constraints, with comparable performance at sampling rates of \unit[100]{Hz} and \unit[15]{Hz} for the tested trajectory.

To further explore the distance limits, we considered $\beta$-angle tracking, for which the coil geometry provides greater torque authority than for motion in the $\alpha$ plane, which is parallel to the coil cover. At the largest demonstrated working distance of \unit[549]{mm} from the coil cover, sinusoidal $\beta$-angle tracking with amplitude $5.5\deg$ is maintained despite an active current constraint (see \Cref{fig:max_distance}), with a relative torque residual of $1.3\times10^{-6}$ and an RMS tracking error of $0.24\deg$.

\section{Control}
\label{sec:Control}

To control the micro-guidewire, we consider two operating modes:
direct torque commands provided by a human operator and automatic
reduced-attitude regulation augmented with a simple repetitive control law. The primary objective of the latter is to establish reproducible experimental conditions for comparing field-alignment and torque-based allocation at matched tip positions and angular trajectories, while demonstrating the achievable tracking accuracy. Repetitive control additionally illustrates the potential to compensate disturbances that recur periodically, motivating future investigation of its applicability to approximately
periodic physiological motion.

We would like to emphasize that pose regulation is not necessarily the most appropriate control strategy for in-vivo navigation. In uncertain,
contact-rich environments, accurate pose tracking alone does not
constrain interaction forces with vessel walls. The choice of
feedback variables therefore remains an important question for
clinical translation. Potential extensions include hybrid pose/force
control and compliance control, with disturbance observers providing
model-based estimates of interaction forces.

\subsection{Manual Torque Control}

For manual navigation, the operator commands the two controllable
torque components using a joystick and adjusts these commands based
on visual observation of the guidewire. No automatic attitude-error
feedback is applied in this mode; nevertheless, the allocation
continuously uses the measured tip position and dipole direction to
realize the commanded effective torque. The operator thus closes
the steering loop, while pose measurements remain essential to
current allocation.

The supplementary video demonstrates navigation through the vascular
model using this mode together with the motorized advancer. Whereas
the fixed-insertion tracking experiments in
\Cref{fig:allocation_limit,fig:max_distance} involve modest angular
deflections imposed by the surrounding vessel geometry, coordinated
advancement and torque commands enable traversal of a larger portion
of the model and require larger bending torques up to \unit[10]{$\mu$Nm}. 

\subsection{Reduced Attitude Feedback Control}

Under the effective-torque model in \Cref{sec:optimization}, both the magnetic torque and the force-induced moment are orthogonal to the dipole axis $\n$. Rotation about this axis is therefore not controllable. Accordingly, we control the reduced attitude $\n\in\mathbb{S}^{2}$, where $\mathbb{S}^{2}$ denotes the unit sphere in $\mathbb{R}^{3}$.

Let $\n_{\SP}\in\mathbb{S}^{2}$ denote the desired dipole direction. We employ a proportional-derivative feedback law previously used in electromagnetic levitation control~\cite{singh2026levitation}. Formulated directly in terms of the dipole's reduced attitude, this law avoids singularities associated with local angular parameterizations:
\begin{align}
   \Bprefix{\bm{\tau}}_{\mathrm{fb}}
    &=
    -\mathbf{K}_{d}\,\Bprefix{\bm{\omega}}
    + k_p\R\left(\n\times\n_{\SP}\right),
    \label{eq:reduced_attitude_feedback}
\end{align}
where $\Bprefix{\bm{\omega}}\in\mathbb{R}^{3}$ denotes the body-fixed angular-velocity, and $k_p>0$, $\mathbf{K}_{d}\succeq0$ are proportional and damping gains, respectively. 

In the water-filled vascular model, the identified damping ratio $D=\sidDWater$ indicates sufficient intrinsic damping for the experiments considered, and we therefore set $\mathbf{K}_{d}=\mathbf{0}$. In air, derivative feedback provides additional damping for the strongly underdamped dynamics identified in \Cref{fig:sysid_beta}. For these experiments, we deploy a Kalman filter that explicitly accounts for measurement delay to mitigate the loss of phase margin associated with delayed feedback. The experiments in air are presented in the supplementary video.

\subsection{Repetitive Control}
\label{sec:repetitive_control}

Periodic reference trajectories and repeatable plant dynamics
produce tracking errors that recur across successive cycles. We
exploit this repeatability by learning a periodic feedforward
correction using a repetitive-control
scheme~\cite{longman2000iterative,zughaibi2021pickPlace}. Let
$j=0,1,\ldots$ denote the iteration index and $i=0,\ldots,M-1$ the
sample index within a cycle. A discretized reference period
contains $M$ samples and has duration $T=MT_s$.

For each angle $\phi\in\{\alpha,\beta\}$, the vector
$\bm{u}_{\mathrm{rc},\phi}\in\mathbb{R}^{M}$ stores the periodic
torque correction. At sample $i$ of cycle $j$, the applied
correction and measured tracking error are
\begin{align}
    \tau_{\mathrm{rc},\phi}^{[j]}[i]
    &=u_{\mathrm{rc},\phi}[i],
    &
    e_{\phi}^{[j]}[i]
    &=\phi_{\SP}^{[j]}[i]-\phi^{[j]}[i].
\end{align}
To compensate for the lag in the closed-loop response, we use a learning update with a phase lead:
\begin{align*}
    u_{\mathrm{rc},\phi}
    [(i-d)\tmod M]
    \leftarrow
    u_{\mathrm{rc},\phi}
    [(i-d)\tmod M]
    +
    k_{\mathrm{rc}}e_{\phi}^{[j]}[i],
\end{align*}
where $k_{\mathrm{rc}}$ is the learning gain and the integer shift $d$ is selected such that $dT_s\approx$ \unit[65]{ms} approximates the manually measured lag in the closed-loop response. Thus, the correction is associated with a phase that precedes the observed
error. This update law effectively introduces integral action in the iteration domain by accumulating tracking errors over successive repetitions.

After each cycle, the updated torque correction is smoothed using a centered moving-average $Q$-filter with an odd window length of $N_q = 17$ samples and wraparound at the cycle boundaries \cite{longman2000iterative}. This symmetric filtering introduces no time delay and attenuates high-frequency components to limit learning near the guidewire resonance. \Cref{fig:allocation_limit,fig:max_distance,fig:fa_vs_torque_summary_075hz} illustrate the repetitive controller in operation, with its output added to the torque command generated by the reduced-attitude controller.

\subsection{Field-Alignment Benchmark}
\label{sec:fa_benchmark}

We compare torque-based allocation with conventional field alignment
for identical angular references and magnet positions. In field-alignment, the resulting magnetic torque is locally equivalent to proportional feedback~\cite{zughaibi2025workspace}. For illustration, consider the magnetic torque about a single axis:
\begin{align*}
    \tau_{\phi}
    &= \mAbs\norm{\b}\sin(u_{\phi}-\phi)
    \approx k_p(u_{\phi}-\phi),
    & k_p &= \mAbs\norm{\b},
\end{align*}
where $u_{\phi}$ denotes the commanded field angle. We therefore
pair each prescribed field magnitude with the corresponding
torque-controller gain to match the local magnetic stiffness.

To obtain comparable angular tracking, we apply the repetitive controller described above to learn a field-direction correction, $u_{\phi}=\phi_{\SP}+u_{\mathrm{rc,fa},\phi}$, while retaining the prescribed field magnitude. The desired field $\b_{\SP}$ is allocated through $\current=\Act_\b(\p)^\dagger\b_{\SP}$, where the allocation matrix depends on position but not on dipole orientation. Feedback is introduced here to match tracking performance and isolate the effect of the allocation strategy on energy consumption. In practice, when dipole-orientation information is available, as required by this feedback controller, it should be used directly for torque-based allocation to exploit its energy-efficiency benefits.

\begin{figure}[t]
    \centering
    \includegraphics[width=0.98\columnwidth]
    {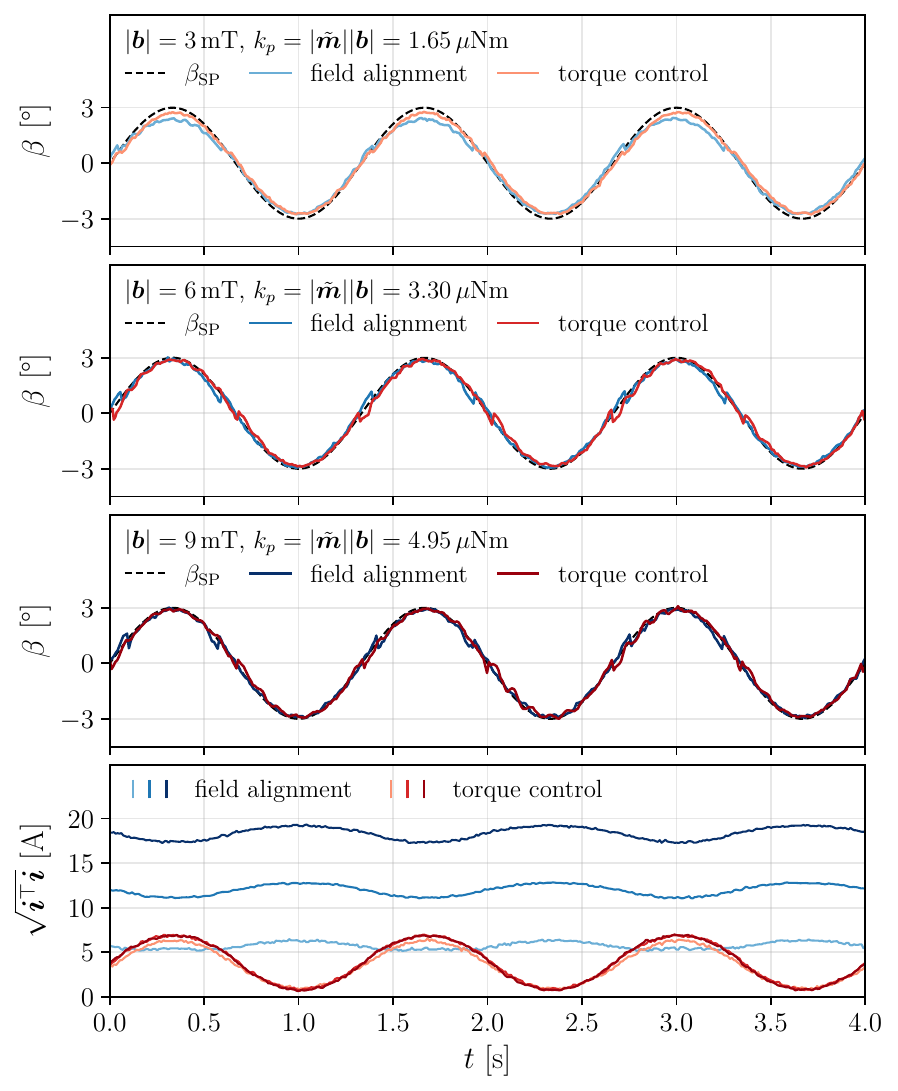}%
    \vspace{\figcapabove}
    \caption{Field alignment versus torque-based control. Top three panels: comparable tip-angle tracking at matched magnetic stiffness $k_p=\mAbs\norm{\b}$. Bottom: current norm $\norm{\current}_2$, showing higher overall current demand under field alignment. The prescribed field cannot be reduced arbitrarily while retaining intuitive manual steering for a human operator. In torque-based control, the field magnitude is an inherent outcome of the current allocation computed to generate the commanded torque. Occasional artifacts in the camera-based measurements are attributed primarily to air bubbles and optical refraction through the vascular model. }
    \label{fig:fa_vs_torque_summary_075hz}
    \vspace{\figcapbelow}
\end{figure}

\Cref{fig:fa_vs_torque_summary_075hz} shows comparable angular tracking with substantially lower current demand under torque-based allocation. In field-alignment, the dominant field component is parallel to the dipole and generates no direct magnetic torque, effectively wasting a significant amout of magnetic energy.

\section{Discussion and Perspective}
\label{sec:Conclusion}

In this work, we demonstrate two complementary benefits of real-time pose information for electromagnetic guidewire control using a clinically oriented three-coil eMNS. First, it enables energy-efficient allocation by exploiting the nonuniqueness of magnetic torque generation, reducing current demand and expanding the usable workspace. Second, it enables dynamic feedback that exploits the high actuation bandwidth of the eMNS. Optimized redistribution of coil currents further maintains accurate torque generation when individual coils reach their current limits. Together with feedback and repetitive control, this enables sub-degree tracking at working distances up to \unit[55]{cm} for a micro-guidewire with a tip magnetic volume of \unit[2.47]{mm$^3$} inside realistic anatomical models.

Clinical translation requires reliable in-vivo state estimation and registration to the eMNS. Retaining the efficiency benefit at \unit[15]{Hz} supports investigating X-ray-based pose feedback, while also motivating complementary non-ionizing localization techniques that can operate in vivo.

These dynamic capabilities motivate exploration of partially automated navigation and automatic rejection of physiological disturbances. How the operator should guide these procedures, which quantities should be controlled, and how the corresponding feedback algorithms should be designed remain open questions for in-vivo applications.




\section*{Acknowledgment}
The authors would like to thank the team of Swissvascular GmbH for the technical support. Further thanks go to the Max Planck ETH Center for Learning Systems for the financial support.

\section*{Conflict of Interests}
Bradley Nelson is the co-founder of MagnebotiX AG and Nanoflex Robotics AG, which commercialize the Navion system. The other authors declare no conflict of interest.

\bibliographystyle{IEEEtran}
\bibliography{bibliography}

\end{document}

%% file: img/sysid_beta_values.tex
\newcommand{\sidKstatAir}{9.8}

\newcommand{\sidfZeroAir}{4.8}
\newcommand{\sidDAir}{0.035}
\newcommand{\sidTdAir}{48}

\newcommand{\sidKstatWater}{8.3}

\newcommand{\sidfZeroWater}{2.9}
\newcommand{\sidDWater}{0.40}
\newcommand{\sidTdWater}{15}